\documentclass[twocolumn,showpacs,preprintnumbers,amsmath,amssymb]{revtex4}

\usepackage{graphicx} 
\usepackage{dcolumn}  
\usepackage{bm}       

\begin{document}

\title{Equation of state of atomic systems
beyond $s$-wave 
determined by the lowest order constrained variational method:
Large scattering length limit}

\author{Ryan M. Kalas$^{(1)}$ and D. Blume$^{(1,2)}$}
\affiliation{%
$^{(1)}$Department of 
Physics and Astronomy, Washington State University, Pullman,
Washington 99164-2814\\
$^{(2)}$INFM-BEC, Dipartimento di Fisica, 
Universit\`a di Trento, I-38050 Povo,
Italy}

\begin{abstract}
Dilute Fermi systems with large $s$-wave scattering length
$a_s$ exhibit  universal properties if the interparticle
spacing $r_o$ greatly exceeds the range of the underlying
two-body interaction potential. In this regime, $r_o$ is
the only relevant length scale and observables such as the
energy per particle depend only on $r_o$ (or,
equivalently, the energy $E_{FG}$ of the free Fermi gas).
This paper investigates Bose and Fermi 
systems with non-vanishing angular momentum $l$
using the lowest order constrained variational method. We focus
on the regime where the generalized scattering length becomes large
and determine the relevant length scales. 
For Bose gases with large generalized scattering lengths,
we obtain simple
expressions for the energy per particle in terms of a 
$l$-dependent length scale $\xi_l$, which depends on the range of the
underlying two-body potential and the average interparticle
spacing.
We discuss possible implications for dilute two-component 
Fermi systems with finite $l$.
Furthermore, 
we determine the equation
of state of 
liquid and gaseous bosonic helium.

\end{abstract}

\maketitle


\section{Introduction}
The experimental realization
of dilute degenerate Bose and Fermi gases has led to an
explosion of activities in the field of cold atom gases.
A particularly intriguing feature of
atomic Bose and Fermi gases is that
their
interaction strengths
can be tuned experimentally through the application
of an external magnetic field
in the vicinity of a Feshbach resonance~\cite{stwa76,ties93}.
This external knob allows dilute systems with essentially any interaction
strength, including infinitely strongly
attractive and repulsive interactions, to be realized.
Feshbach resonances have been experimentally observed for
$s$-, $p$- and $d$-wave interacting
gases~\cite{inou98,cour98,rega03,zhan04,chin00}
and have been predicted
to exist also for higher partial waves.

A Feshbach resonance arises due to the coupling of two
Born-Oppenheimer potential curves coupled through a
hyperfine Hamiltonian, and requires, in general, a
multi-channel description. For $s$-wave interacting systems,
Feshbach resonances can be classified as broad or narrow~\cite{kohl06}.
Whether a resonance is broad or narrow depends on whether the energy
width of the resonance is large or small compared to the characteristic energy
scale, such as the Fermi energy or the harmonic oscillator
energy, of the system. 
In contrast to $s$-wave resonances,
higher partial wave resonances are necessarily narrow
due to the presence of the angular momentum barrier~\cite{land77}.
This paper uses an effective single channel description
to investigate the behaviors of strongly-interacting Bose and
Fermi systems with different orbital angular momenta.

In dilute homogeneous Bose and Fermi gases with
large $s$-wave scattering length $a_s$, a regime
has been identified in which the energy per particle takes on a universal
value which is set by a single length scale,
the average interparticle spacing 
$r_o$~\cite{bake99,heis01,cowe02}.
In this so-called unitary regime, the length scales
of the $s$-wave interacting system
separate according to $|a_s| \gg r_o \gg R $, where
$R$ denotes the range of the two-body potential.
The energy per particle $E_{B,0}/N$ (the subscripts ``$B$'' and ``$0$''
stand respectively for
``boson'' and ``$s$-wave interacting'') for a homogeneous one-component
gas of bosons
with mass $m$ in the unitary regime
has been calculated to be
$E_{B,0}/N \approx 13.3 \:\hbar^2 n_B^{2/3}/m$
using the lowest order constrained variational
(LOCV) method~\cite{cowe02}. 
The energy $E_{B,0}/N$ at unitarity is thus independent
of $a_s$ and $R$, and depends on the single length
scale $r_o$ through
the boson number density $n_B$, $r_o=(4 \pi n_B / 3)^{-1/3}$.
However, Bose gases in the
large scattering length limit are expected to be unstable
due to three-body recombination~\cite{fedi96,esry99b,niel99,beda00}.

On the other hand, the Fermi pressure prevents the collapse of
two-component Fermi gases with equal masses and equal number of ``spin-up''
and ``spin-down''
fermions with
large interspecies $s$-wave scattering 
length~\cite{bake99,heis01,hara02,bour03}.
At unitarity, the energy per particle is
given by $E_{F,0}/N \approx 0.42 E_{FG}$, where
$E_{FG} = (3/10)(\hbar^2 k_F^2 / m)$ denotes the energy per particle of the
non-interacting Fermi gas~\cite{carl03,chan04,astr04c,carl05}.
The Fermi wave vector $k_F$
is related to the number density of the
Fermi gas by $n_F = k_F^3/3\pi^2$, which implies that $E_{F,0}/N$
depends on $r_o$ but is independent of $a_s$ and $R$.
We note that the inequality $|a_s|\gg r_o$
is equivalent to $1/(k_F|a_s|) \ll 1$.

This paper investigates
Bose and Fermi systems with large generalized scattering lengths
using the LOCV method.
For $p$- and $d$-wave interacting Bose systems, we define
the unitary regime~\cite{footnote3} 
through the inequalities $|a_l(E_{rel})| \gg \xi_l \gg R$,
where $\xi_l$ denotes a $l$-dependent length scale
given by the geometric
combination of $r_o$ and $R$, i.e.,
$\xi_l = r_o^{(1-l/4)} R^{l/4}$, and $E_{rel}$ the relative
scattering energy.
The generalized energy-dependent
scattering length $a_l(E_{rel})$~\cite{blum02,bold02,stoc04}
characterizes the scattering
strength (see below).
We find that the energy of $p$-wave interacting two-component
Bose gases and $d$-wave
interacting one- and two-component
Bose gases at unitary is determined by the combined
length $\xi_l$.
While Bose gases with higher angular momentum
in the unitary regime are
of theoretical interest, they
are, like their $s$-wave cousin, expected to be unstable.
We comment that the energetics of
two-component Fermi gases with large generalized
scattering length may depend on the same length scales.

Furthermore, 
we consider $s$-wave interacting Bose systems over a wide range
of densities. Motivated by two recent studies by Gao~\cite{gao04,gao05},
we determine the energy per particle $E_{B,0}/N$ 
of the Bose system characterized
by two atomic physics parameters, the $s$-wave scattering lengh $a_s$
and the van der Waals coefficient $C_6$. 
Our results lead to a phase diagram of liquid helium
in the low-density regime that 
differs from
that proposed in Ref.~\cite{gao05}.

Section~\ref{sectionII} describes the systems under study
and introduces the LOCV method. Section~\ref{sectionIII}
describes our results for dilute $s$-wave interacting
Bose and Fermi systems 
and for liquid helium.  Section~\ref{sectionIV}
considers 
Bose and Fermi systems interacting through $l$-wave ($l>0$)
scattering.  Finally, Section~\ref{sectionV} concludes.

\section{LOCV method for bosons and fermions}
\label{sectionII}
This section introduces the three-dimensional
Bose and Fermi systems under study
and reviews the LOCV method~\cite{pand71,pand71a,pand73,pand77}.
The idea of the LOCV method is to explicitly treat 
two-body correlations, but to neglect three- and higher-body
correlations.
This allows the many-body problem to be reduced to solving
an effective two-body equation
with properly chosen constraints.
Imposing these constraints makes the method non-variational, i.e.,
the resulting energy does not place an upper bound on the
many-body energy.
The LOCV method is expected to capture some of the key physics
of dilute Bose and Fermi systems.

The Hamiltonian $H_B$
for a homogeneous system consisting of identical mass $m$
bosons is given by
\begin{equation}
\label{hb}
H_B =  -\frac{\hbar^2}{2m}\sum_i\nabla^2_i+\sum_{i<j}v(r_{ij}),
\end{equation}
where
the spherically symmetric interaction potential $v$
depends on the
relative
distance $r_{ij}$, $r_{ij}=|{\bf r}_i - {\bf r}_j|$.
Here,
${\bf{r}}_i$ denotes the position vector of the $i$th boson.
The Hamiltonian $H_F$ for a two-component Fermi system with equal masses
and identical spin population is given by
\begin{equation}
\label{hf}
H_F =  -\frac{\hbar^2}{2m}\sum_{i}\nabla^2_i
-\frac{\hbar^2}{2m}\sum_{i'}\nabla^2_{i'}+
\sum_{i,i'}v(r_{ii'}),
\end{equation}
where the unprimed subscripts label spin-up
and the primed subscripts spin-down fermions.
Throughout,
we take like fermions to be non-interacting.
Our primary interest in this paper is in the description of 
systems for which many-body observables are insensitive
to the short-range behavior of the atom-atom potential $v(r)$.
This motivates us to consider
two simple model potentials:
an attractive square well potential $v_{sw}$ with depth $V_o$ ($V_o \ge 0$),
\begin{equation}
\label{vsw}
v_{sw}(r) = \left\{
\begin{array}{c l}
  -V_o  &  \mbox{ for } \quad  r<R \,  \\
  0     &  \mbox{ for } \quad  r>R \, ;
\end{array}
\right.
\end{equation}
and an attractive van der Waals potential $v_{vdw}$ with hardcore $r_c$,
\begin{equation}
v_{vdw}(r) = \left\{
\begin{array}{c l}
  \infty       &  \mbox{ for } \quad  r<r_c \,   \\
  -C_6/r^6     &  \mbox{ for } \quad  r>r_c \, .
\end{array}
\right.
\end{equation}
In all applications, we choose the hardcore $r_c$
so that the inequality $r_c\ll\beta_6$, where  $\beta_6 = (mC_6/\hbar^2)^{1/4}$,
is satisfied.
The natural length scale of the square well potential is given by
the range $R$ and that of the van der Waals potential by
the van der Waals length $\beta_6$.
The solutions to the
two-body Schr\"odinger equation for $v_{sw}$ are given in
terms of spherical Bessel
and Neumann functions (imposing the proper continuity conditions 
of the wave function and its derivations at those $r$
values where the potential exhibits a discontinuity), 
and those for $v_{vdw}$ in terms
of convergent infinite series of spherical
Bessel and Neumann functions~\cite{gao98}.

The interaction strength of the short-range square well
potential can be characterized
by the generalized energy-dependent scattering 
lengths $a_l(k)$,
\begin{eqnarray}
\label{eq_scatt1}
a_l(k)= \mbox{sgn}[- \tan \delta_l(k)]
\left| \frac{\tan\delta_l(k)}{k^{2l+1}} \right|^{1/(2l+1)},
\end{eqnarray}
where $\delta_l(k)$ denotes the phase shift of the $l$th partial wave
calculated at the relative scattering energy $E_{rel}$,
$k=\sqrt{m E_{rel}/\hbar^2}$.
This definition ensures that
$a_l(k)$ approaches a constant as $k \rightarrow 0$~\cite{mott65,tayl72}.
For the van der Waals potential $v_{vdw}$, the threshold behavior changes for
higher partial waves and the
definition of $a_l(k)$ has to be modified
accordingly~\cite{mott65,tayl72}.
In general,
for a potential that falls off as $-r^{-n}$ at large interparticle
distances, $a_l(k)$ is defined by Eq.~(\ref{eq_scatt1})
if $2l < n-3$ and by
\begin{eqnarray}
\label{eq_scatt2}
a_l(k)= \mbox{sgn}[- \tan \delta_l(k)] \left|
\frac{ \tan\delta_l(k)}{k^{n-2}} \right|^{1/(n-2)}
\end{eqnarray}
if $2l>n-3$.
For our van der Waals potential, $n$ is equal to $6$ and $a_l(k)$ is given by
Eq.~(\ref{eq_scatt1})
for $l \le 1$ and by Eq.~(\ref{eq_scatt2}) for $l \ge 2$.
The zero-energy generalized scattering lengths
$a_l$ can now be defined readily through
\begin{eqnarray}
a_l = \lim_{k \rightarrow 0} a_l(k).
\end{eqnarray}

We note that a new two-body $l$-wave bound state appears at threshold
when $|a_l| \rightarrow \infty$. The unitary regime for
higher partial waves discussed in Sec.~\ref{sectionIV}
is thus, as in the $s$-wave case, closely related to the physics
of extremely weakly-bound atom-pairs.
To uncover the key behaviors at unitarity, we
assume in the following that the
many-body system under study is interacting through a single partial
wave $l$.
While this may not be exactly realized in an experiment,
this situation may be approximated by utilizing
Feshbach resonances.

We now outline how the energy per particle $E_{B,l}/N$ of 
a one-component Bose system
with $l$-wave interactions~\cite{footnote5}
is calculated by the LOCV method~\cite{pand71,pand71a,pand73,pand77}.
The boson wave function $\Psi_B$
is taken to be a product of pair functions $f_l$,
\begin{equation}
\label{psib}
\Psi_B({\bf r}_1,\ldots,{\bf r}_N)=\prod_{i<j}f_l(r_{ij}),
\end{equation}
and the energy expectation value of $H_B$, Eq.~(\ref{hb}),
is calculated using $\Psi_B$.
If terms depending on the coordinates of three or more
different particles are
neglected,
the resulting energy is given by
the two-body term in the cluster expansion,
\begin{equation}
\label{ebcluster}
\frac{E_{B,l}}{N}=\frac{n_B}{2}
\int f_l(r) \Big{[} -\frac{\hbar^2}{m}\nabla^2 +v(r)
\Big{]} f_l(r)\, {\rm d}^3{\bf r}.
\end{equation}
The idea of the LOCV method is now to introduce a healing distance $d$
beyond which the pair correlation function $f_l$ is constant,
\begin{eqnarray}
\label{fdfdr_1}
f_l(r>d)=1.
\end{eqnarray}
To ensure that the derivative of $f_l$
is continuous at $r=d$,
an additional constraint is introduced,
\begin{eqnarray}
\label{fdfdr_2}
f'_l(r=d)=0.
\end{eqnarray}
Introducing a constant average
field $\lambda_l$ and varying with respect
to $f_l$ while using that $f_l$ is constant for $r>d$,
gives the Schr\"odinger-like two-body equation for $r<d$,
\begin{equation}
\label{schro}
\Big{[} -\frac{\hbar^2}{m}\nabla^2 + v(r) \Big{]}\left(rf_l(r)\right)=
\lambda_l r f_l(r).
\end{equation}
Finally, the condition
\begin{equation}
\label{norm1}
n_B \int_0^d \! f_l^2(r)\; {\rm d}^3{\bf r} =1
\end{equation}
enforces that the average number of particles within
$d$ equals $1$.
Using Eqs.~(\ref{ebcluster}), (\ref{fdfdr_1}) and (\ref{schro}),
the energy per particle becomes,
\begin{equation}
\label{ebn}
\frac{E_{B,l}}{N} = 
\frac{\lambda_l}{2} + \frac{n_B}{2} \int_d^\infty v(r) {\rm d}^3{\bf r}.
\end{equation}
The
second term on the right hand side of Eq.~(\ref{ebn})
is identically zero for the square well potential $v_{sw}$
but contributes a so-called
tail or mean-field energy for the van der Waals potential
$v_{vdw}$~\cite{gao04,gao05}.
We determine the three unknown $n_B$, $\lambda_l$ and $d$
by simultaneously solving
Eqs.~(\ref{schro}) and (\ref{norm1}) subject to the boundary condition
given by Eq.~(\ref{fdfdr_2}). Note that $n_B$ and $d$ depend,
just as $f_l$ and $\lambda_l$, on the angular momentum;
the subscript has been dropped, however, for notational convenience.

In addition to one-component Bose systems, Sec.~\ref{sectionIV}
considers two-component Bose systems,
characterized by $l$-wave interspecies and 
vanishing intraspecies interactions. The Hamiltonian for
the two-component Bose
system is given by Eq.~(\ref{hf}), with the
sum of the two-body interactions restricted to 
unlike bosons. Correspondingly, the product
wave function is written as a product of pair
functions, including only correlations
between unlike bosons.
The LOCV equations are then given by Eqs.~(\ref{fdfdr_1})
through (\ref{norm1}) with $n_B$ in Eq.~(\ref{norm1}) replaced
by $n_B/2$.

Next, we discuss how to determine the energy $E_{F,l}/N$ per particle
for a two-component Fermi system
within the LOCV method~\cite{pand71,pand71a,pand73,pand77}.
The wavefunction is taken to be
\begin{equation}
\label{eq_wavef}
\Psi_F({\bf r}_1,\ldots,{\bf r}_{1'},\dots)=\Phi_{FG}\prod_{i,j'}f_l(r_{ij'}),
\end{equation}
where $\Phi_{FG}$ denotes the ground state wavefunction of the
non-interacting Fermi gas. The product of
pair functions $f_l$ accounts for the correlations between
unlike fermions.
In accord with our assumption that like fermions are non-interacting,
Eq.~(\ref{eq_wavef})
treats like fermion pairs as uncorrelated.
Neglecting exchange effects,
the derivation of the LOCV equations parallels that outlined above
for the bosons.
The boundary conditions,
given by Eqs.~(\ref{fdfdr_1}) and (\ref{fdfdr_2}),
and the Schr\"odinger-like differential equation
for $\lambda_l $, Eq.~(\ref{schro}), are unchanged.
The ``normalization condition,'' however, becomes
\begin{equation}
\label{norm2}
\frac{n_F}{2} \int_0^d f_l^2(r){\rm d}^3{\bf r} =1,
\end{equation}
where the left-hand side
is the number of fermion pairs within $d$.
The fermion energy per particle is then the sum of
the one-particle contribution from the
non-interacting Fermi gas and the pair correlation energy 
$\lambda_l$~\cite{chan04},
\begin{equation}
\label{efn}
E_{F,l}/N = E_{FG} + \frac{\lambda_l}{2}.
\end{equation}
This equation excludes the contribution from the tail
of the potential, i.e., the
term analogous to the second term on the right hand side of
Eq.~(\ref{ebn}), since this term is negligible
for the fermion densities considered in this paper.

The LOCV solutions for $f_l$, $\lambda_l$ and $d$ for the homogeneous 
one-component Bose system
and the two-component Fermi system are
formally identical if the boson density is
chosen to equal half the fermion density, i.e.,
if $n_B = n_F/2$.  This relation can be understood by realizing that
any given fermion (e.g., a spin-up particle) interacts
with only half of the total number of fermions
(e.g., all the spin-down fermions). Consequently,
the two-component Fermi system appears twice as
dense as the one-component Bose system.
The fact that the LOCV solutions for bosons can be converted to
LOCV solutions for fermions suggests that some physics of the bosonic system
can be understood in terms of the fermionic system and vice
versa.  In fact, it has been shown previously~\cite{chan04}
that the LOCV energy for the first excited gas-like state
of $s$-wave interacting fermions at unitarity
can be derived from the LOCV energy of the
energetically lowest-lying gas-like branch of $s$-wave interacting 
bosons~\cite{cowe02}.
Here, we extend this analysis and show that the ground state
energy of the Fermi gas at unitarity can be
derived from the energetically highest-lying liquid-like branch of
the Bose system. Furthermore, we extend this
analysis to
higher angular momentum scattering.

\section{$s$-wave interacting Bose and Fermi systems}
\label{sectionIII}
\begin{figure}
\vspace{0.2in}
\includegraphics[scale=0.5]{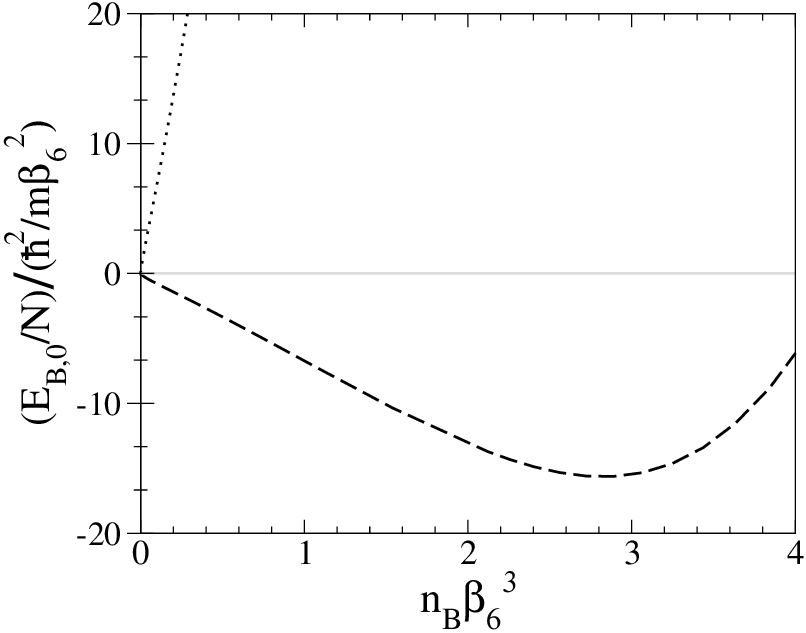}
\caption{\label{cap1}
Energy per particle $E_{B,0}/N$ as a function of the density $n_B$,
both plotted as dimensionless quantities, for a one-component 
Bose system
interacting through the van
der Waals potential with $s$-wave
scattering length $a_s=16.9 \beta_6$.
The dotted line shows the gas branch and the
dashed line the liquid branch.  The minimum of the liquid branch is discussed
in reference to liquid $^4$He in the text.
}
\end{figure}

Figure~\ref{cap1} shows the energy per particle
$E_{B,0}/N$, Eq.~(\ref{ebn}),
obtained by solving the LOCV equations for a 
one-component Bose system
interacting through the van der Waals
potential with
$s$-wave scattering length $a_s=16.9 \beta_6$.
The dotted line in Fig.~\ref{cap1} has positive energy and
increases with increasing density; it describes
the energetically lowest-lying ``gas branch'' for the Bose system
with $a_s=16.9 \beta_6$ and corresponds to the metastable gaseous condensate
studied experimentally.
The dashed line in Fig.~\ref{cap1} has negative energy at small densities,
decreases with increasing density, and then
exhibits a minimum; this dashed line describes the energetically
highest-lying ``liquid branch''
for a Bose system
with $a_s=16.9 \beta_6$.
Within the LOCV framework, these two branches
arise because the Schr\"odinger-like equation, Eq.~(\ref{schro}),
permits for a given interaction potential
solutions $f_0$ with differing number of nodes,
which in turn give rise to
a host of liquid and
gas branches~\cite{gao04,gao05}.
Throughout this work we only consider the energetically
highest-lying liquid branch
with $n$ nodes and the
energetically lowest-lying gas branch with $n+1$ nodes.
To obtain Fig.~\ref{cap1}, we consider a class of two-body potentials
with fixed $a_s/\beta_6$,
and decrease the value of the ratio $r_c/\beta_6$ till $E_{B,0}/N$,
Eq.~(\ref{ebn}), no longer changes over the density
range of interest, i.e.,
the number of nodes $n$ of the energetically highest-lying 
liquid branch is increased till
convergence is reached.

In Fig.~\ref{cap1}, the two-body van der Waals potential is chosen
so that
the scattering length of $a_s=16.9 \beta_6$ coincides with
that of the $^4$He pair potential~\cite{janz95}.
The liquid branch in Fig.~\ref{cap1} can hence be applied to
liquid $^4$He, and has previously been considered in 
Refs.~\cite{gao04,gao05}.
The minimum of the liquid branch at a density of $n_B=2.83 \beta_6^{-3}$,
or $1.82\times 10^{22} {\rm cm}^{-3}$,
agrees quite well with
the experimental
value of $2.18\times 10^{22}{\rm cm}^{-3}$~\cite{oubo87}.
The corresponding energy per particle of $-6.56$~K
deviates by 8.5~\% from the experimental value of 
$-7.17$~K~\cite{oubo87}.
This shows that the LOCV framework provides a fair description of the strongly
interacting liquid $^4$He system, which is characterized by interparticle
spacings comparable to the range of the potential.
This is somewhat remarkable considering that the LOCV method
includes only pair correlations and that the van der Waals
potential used here contains
only two parameters.

\begin{figure}
\vspace{0.2in}
\includegraphics[scale=0.5,angle=0]{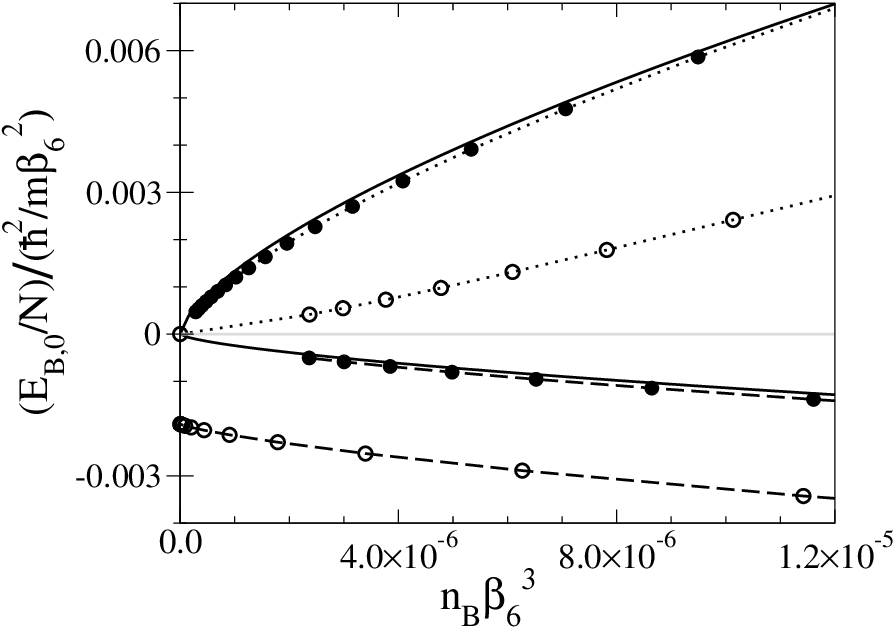}
\caption{\label{cap2}
Energy per particle $E_{B,0}/N$ as a function of the density $n_B$
for a one-component Bose system
interacting through the van der Waals potential with $s$-wave
scattering lengths $a_s=16.9 \beta_6$ (open circles) and
$a_s=169 \beta_6$ (filled circles).
To guide the eye, dashed and dotted lines
connect the data points of the liquid and gas branches, respectively.
The liquid branches go to
$E_{dimer}/2$ as the density goes to zero.
The solid lines show $E_{B,0}/N$ at unitarity; see text for discussion.
Compared to Fig.~\protect\ref{cap1},
the energy and density scales are greatly enlarged.
}
\end{figure}

Open circles connected by a
dashed line in Fig.~\ref{cap2} show the liquid branch for $a_s=16.9 \beta_6$
in
the small density region.
As the
density goes to zero,
the energy per particle $E_{B,0}/N$
does not terminate at zero but, instead,
goes to $E_{dimer}/2$, where $E_{dimer}$ denotes the
energy of the most weakly-bound $s$-wave molecule of $v_{vdw}$.
In this
small density limit,
the liquid branch describes a gas of weakly-bound molecules,
in which the interparticle spacing between the molecules greatly exceeds
the size of the molecules,
and $E_{dimer}$ is to a very good approximation given
by $-\hbar^2/(m a_s^2)$.
As seen in Fig.~\ref{cap2}, we find solutions
in the whole density range considered.
In contrast to our findings,
Ref.~\cite{gao05} reports that
the LOCV solutions of the liquid branch disappear
at densities smaller than a scattering length dependent critical density,
i.e., at a critical density of $8.68 \times 10^{-7} \beta_6^{-3}$
for $a_s=16.9 \beta_6$.
Thus we are not able to reproduce the
liquid-gas phase diagram proposed in Fig.~2
of Ref.~\cite{gao05},
which depends on this termination of the liquid branch.
We note that the liquid branch is, as indicated by its 
imaginary speed of sound, dynamically unstable at sufficiently 
small densities.
The liquid of weakly-bound bosonic molecules discussed
here can, as we show below, be related to weakly-bound molecules
on the BEC side of the BEC-BCS crossover curve for two-component Fermi
gases.

We now discuss the gas branch in more detail.
Open and filled circles connected by dotted lines in Fig.~\ref{cap2}
show the energy per particle for
$a_s=16.9 \beta_6$ and $169 \beta_6$, respectively.
These curves can be applied, e.g., to
$^{85}$Rb, whose scattering length can be tuned by
means of a Feshbach resonance and which has
a $\beta_6$ value of $164 a_{bohr}$, where
$a_{bohr}$ denotes the Bohr radius.
For this system,
a scattering length of $a_s= 16.9 \beta_6$ corresponds to $2770 a_{bohr}$,
a comparatively large value that can be realized experimentally
in $^{85}$Rb gases.
As a point of reference, 
a density of $10^{-5}\beta_6^{-3}$ corresponds to a density of
$1.53\times 10^{13}{\rm cm}^{-3}$ for $^{85}$Rb.

\begin{figure}
\vspace{0.2in}
\includegraphics[scale=0.5]{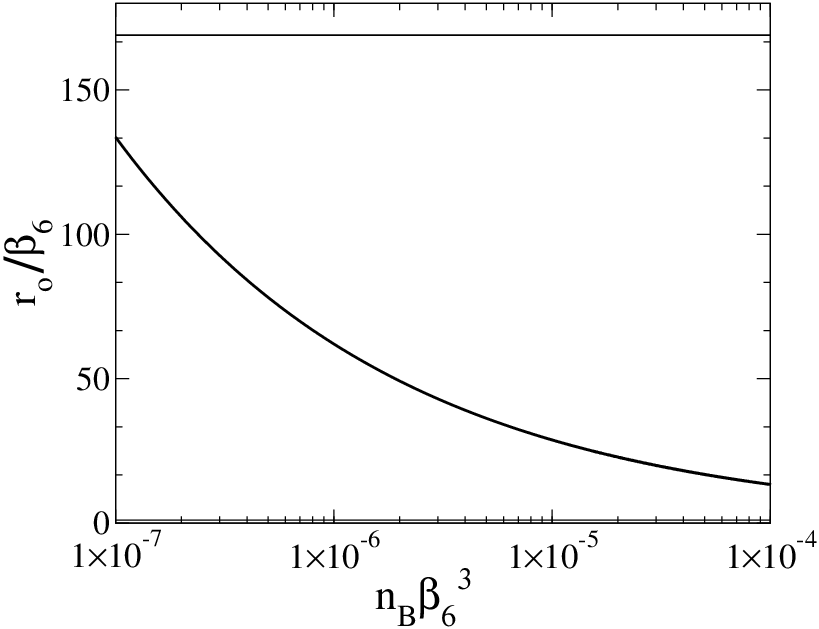}
\caption{\label{cap3}
Scaled interparticle spacing $r_o / \beta_6$ as a function of
the scaled density $n_B \beta_6^3$ for the gas branch of a 
one-component Bose system
interacting through the van der Waals potential
with $a_s=169 \beta_6$.
The horizontal lines show the
scaled $s$-wave scattering length
$a_s = 169 \beta_6$ and the range of the van der Waals
potential, which is one in scaled units
(almost indistinguishable from the $x$-axis). This graph shows
that the unitary inequalities $a_s \gg r_o \gg \beta_6$
hold for $n_B$ larger than about $10^{-5} \beta_6^{-3}$.
}
\end{figure}

The solid curve with positive energy in Fig.~\ref{cap2}
shows the energy per particle $E_{B,0}/N$ at unitarity,
$E_{B,0}/N \approx 13.3 \hbar^2 n_B^{2/3} /m$~\cite{cowe02}.
As seen in Fig.~\ref{cap2}, this unitary limit is approached by
the energy per particle for the Bose gas with $a_s=169 \beta_6$
(filled circles connected by a dotted line).
To illustrate this point,
Fig.~\ref{cap3} shows the scaled average interparticle
spacing $r_o/ \beta_6$ as a function of the scaled density $n_B \beta_6^3$
for $a_s=169 \beta_6$. This plot indicates that
the unitary requirement, $a_s\gg r_o \gg R$, is met
for values of $n_B \beta_6^3$ larger than about $10^{-5}$.
Similarly,
we find that the family of liquid curves converges to
$E_{B,0}/N \approx - 2.46 \hbar^2 n_B^{2/3}/m$ (see Sec.~\ref{sectionIV}
for details),
plotted as a solid line in Fig.~\ref{cap2},
when the inequalities $a_s \gg r_o \gg \beta_6$
are fullfilled.
We note that the unitarity curve with negative energy is also approached,
from above,
for systems with large negative scattering lengths
(not shown in Fig.~\ref{cap2}).
Aside from the proportionality constant, the power law relation
for the liquid and gas branches at unitarity is the same.

In addition to a Bose system interacting through
the van der Waals potential, we consider a
Bose system interacting through the square well
potential with range $R$.
For a given
scattering length $a_s$ and density $n_B$, the energy per particle $E_{B,0}/N$
for these two two-body potentials is
essentially identical for the densities shown in Fig.~\ref{cap2}.
This agreement emphasizes that the details of the two-body potential
become negligible at low density, and in particular, that the behavior
of the Bose gas in the
unitary limit is governed by a single length scale, the average
interparticle spacing $r_o$.

\begin{figure}
\vspace{0.2in}
\includegraphics[scale=0.5]{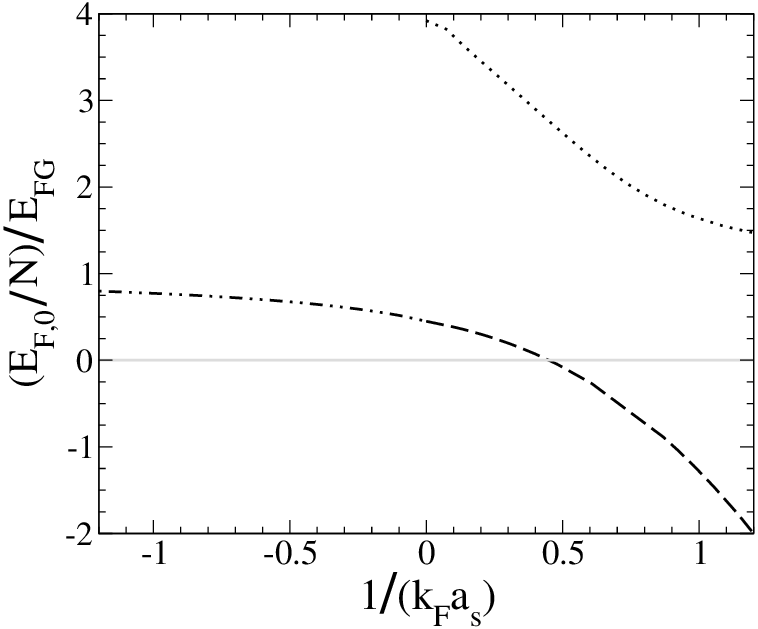}
\caption{\label{cap4}
Scaled energy per particle $(E_{F,0}/N)/E_{FG}$ as a function of
$1/(k_F a_s)$ for a two-component $s$-wave Fermi gas
interacting through the square well potential for
$n_F=10^{-6} R^{-3}$.
The combined dashed and dash-dotted curve
corresponds to the BEC-BCS crossover curve and
the dotted curve corresponds to the first excited state of the Fermi gas.
The dashed and dotted linestyles are chosen
to emphasize the connection to the gas and liquid branches of the Bose
system in Figs.~\protect\ref{cap2}
and \protect\ref{cap3} (see text for more details).
}
\end{figure}

As discussed in Sec.~\ref{sectionII},
the formal parallels between the LOCV method applied to bosons
and fermions allows the energy per particle $E_{F,0}/N$ for a two-component
Fermi gas, Eq.~(\ref{efn}), to be obtained straightforwardly
from the energy per particle $E_{B,0}/N$ of the Bose system.
Figure~\ref{cap4} shows the dimensionless energy
$(E_{F,0}/N)/E_{FG}$ as a function of the dimensionless
quantity $1/(k_F a_s)$
for the square well potential for $n_F=10^{-6}R^{-3}$.
We find essentially identical results for the van der Waals potential.
The
crossover curve shown in Fig.~\ref{cap4} describes
any dilute Fermi gas for which the range $R$ of the two-body potential
is very small compared to the average interparticle spacing $r_o$.
In converting the energies for the Bose system to those for the Fermi
system, the gas branches of the Bose system (dotted lines in
Figs.~\ref{cap2} and \ref{cap3})
``turn into'' the excited state of the Fermi gas
(dotted line in Fig.~\ref{cap4});
the liquid branches of the Bose system with positive $a_s$
(dashed lines in Figs.~\ref{cap2} and \ref{cap3})
``turn into'' the part of the BEC-BCS crossover curve
with positive $a_s$ (dashed line in Fig.~\ref{cap4}); and
the liquid branches of the Bose system with negative
$a_s$ (not shown in Figs.~\ref{cap2} and \ref{cap3}) ``turn into''
the part of the BEC-BCS crossover curve
with negative $a_s$ (dash-dotted line in Fig.~\ref{cap4}).

To emphasize the connection between the Bose and Fermi
systems further, let us consider the
BEC side of the crossover curve.
If $1/(k_F a_s) \gtrsim 1$, the fermion energy per particle $E_{F,0}/N$
is approximately given by $E_{dimer}/2$, which indicates that
the Fermi gas forms a molecular Bose gas.
Similarly, the liquid branch of the
Bose system with positive scattering length is made up of bosonic
molecules as the density
goes to zero.
The formal analogy between the Bose and Fermi LOCV
solutions also
allows the energy per particle $E_{F,0}/N$ at unitarity,
i.e., in the $1/(k_F |a_s|) \rightarrow 0$ limit,
to be calculated from the energies for large $a_s$
of the gas and liquid branches of the
Bose system (solid lines in Fig.~\ref{cap2}).
For the excited state of the Fermi gas
we find $E_{F,0}/N \approx 3.92 E_{FG}$,
and for the lowest gas state
we find $E_{F,0}/N \approx 0.46 E_{FG}$.
These results agree with the LOCV calculations of Ref.~\cite{chan04},
which use an attractive $\cosh$-potential and a $\delta$-function
potential. The value
of $0.46 E_{FG}$ is in good agreement with the energy
of $0.42 E_{FG}$ obtained by
fixed-node diffusion Monte Carlo calculations~\cite{astr04c,carl05}.

\section{Bose and Fermi systems beyond $s$-wave at unitarity}
\label{sectionIV}
\begin{figure}
\vspace{0.2in}
\includegraphics[scale=0.55]{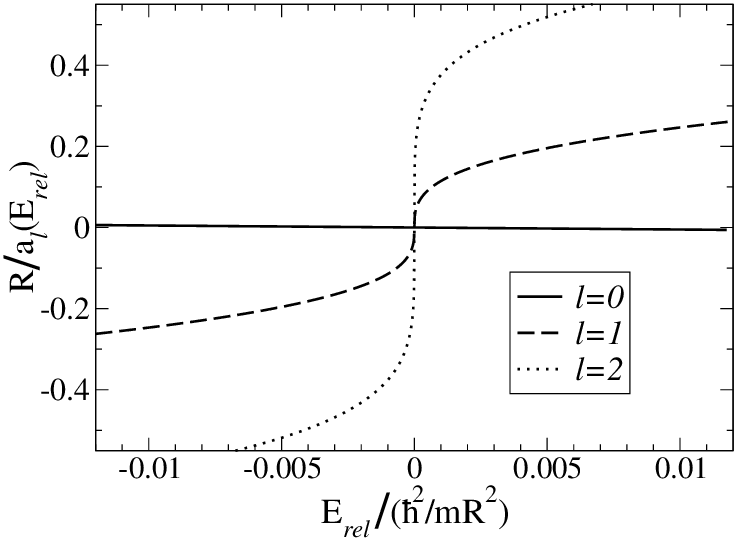}
\caption{\label{cap5} $R/a_l(E_{rel})$ as a function of the scaled relative 
scattering energy $E_{rel}/(\hbar^2/mR^2)$ for
the square well potential $v_{sw}$ with infinite zero-energy
scattering length $a_l$, i.e., $1/a_l=0$,
for three different 
partial waves [$l=0$ (solid line), $l=1$ (dashed line), and
$l=2$ (dotted line)].}
\end{figure}

This section investigates the unitary regime of Bose and
Fermi systems interacting through higher angular momentum resonances.
These higher angular momentum resonances are necessarily
narrow~\cite{land77}, and we hence expect the energy-dependence of
the generalized scattering length $a_l(k)$ to be particularly important
in understanding the many-body physics of dilute atomic systems 
beyond $s$-wave.
In the following we focus on the strongly-interacting limit.
Figure~\ref{cap5} shows $1/a_l(k)$ as a function of the relative 
scattering energy $E_{rel}$ for the square-well potential 
with infinite zero-energy scattering length $a_l$ for
three different angular momenta, $l=0$ (solid line),
$l=1$ (dashed line), and $l=2$ (dotted line).
Figure~\ref{cap5} shows that the 
energy-dependence of $a_l(k)$ increases with increasing
$l$.

Our goal is to determine the energy per particle $E_{B,l}/N$ for 
Bose systems with finite angular momentum $l$ in the strongly-interacting 
regime. For $s$-wave interactions, the only relevant length scale
at unitarity 
is the average interparticle
spacing $r_o$ (see Sec.~\ref{sectionIII}). In this case,
the energy per particle
at unitarity can be estimated analytically
by evaluating the LOCV equations subject to the boundary condition
implied by the zero-range $s$-wave 
pseudo-potential~\cite{cowe02}. Unfortunatey, a similarly simple analysis 
that uses the boundary
condition implied by the two-body
zero-range pseudo-potential for higher partial waves
fails.
This combined with the following arguments suggests
that $E_{B,l}/N$ depends 
additionally on 
the range of the underlying two-body potential for finite $l$:
i) The probability distribution of the two-body $l$-wave bound state,
$l>0$, 
remains finite as $a_l$ approaches infinity
and depends on the interaction potential~\cite{riis92,jens04}.
ii) A description of $l$-wave resonances ($l>0$)
that uses a coupled channel square well model depends
on the range of the square well potential~\cite{gree05}. 
iii) The calculation of structural expectation values of two-body systems
with finite $l$ within
a zero-range pseudo-potential treatment 
requires a new length scale to be introduced~\cite{stoc04}.

Motivated by these two-body arguments 
(see also Refs.~\cite{ho05,chen05,iski06} for
a treatment of $p$-wave interacting Fermi gases)
we propose
the following functional form for the energy per particle $E_{B,l}/N$
of a $l$-wave Bose
system at unitarity 
interacting through the square-well potential $v_{sw}$ with range $R$,
\begin{eqnarray}
\label{ebpower}
\frac{E_{B,l}}{N}=C_l\;\frac{\hbar^2}{mR^{x_l/2}}\;n_B^{2/3-x_l/6}.
\end{eqnarray}
Here, $C_l$ denotes a dimensionless $l$-dependent proportionality constant.
The dimensionless parameter $x_l$ determines the
powers of the range $R$ and the density $n_B$, and ensures
the correct units of the right hand side of Eq.~(\ref{ebpower}).
To test the validity of Eq.~(\ref{ebpower}),
we solve
the LOCV equations, Eqs.~(\ref{fdfdr_2}) through
(\ref{ebn}), for $l=0$ to $2$
for the one-component Bose
system.
Note that the one-component 
$p$-wave system is unphysical since it
does not obey Bose symmetry;
we nevertheless consider it here since its
LOCV energy determines the energy of two-component
$p$-wave Bose and Fermi systems (see below).

Figure~\ref{cap7a}
\begin{figure}
\vspace{0.2in}
\includegraphics[scale=0.55]{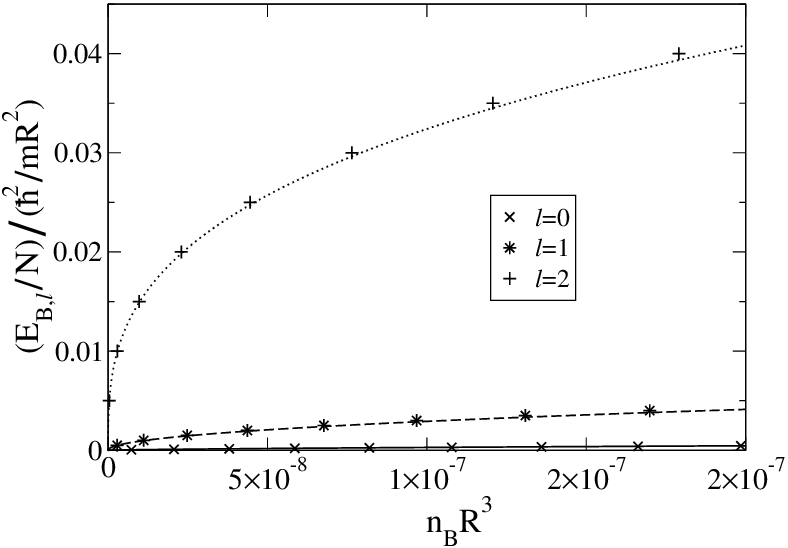}
\caption{\label{cap7a} 
Scaled energy per particle $(E_{B,l}/N)/ (\hbar^2/mR^2)$ for
a one-component Bose system
for the energetically lowest-lying 
gas branch as a function of the scaled
density $n_B R^{3}$ obtained by solving the LOCV equations
[Eqs.~(\protect\ref{fdfdr_2}) 
through 
(\protect\ref{ebn})] for $v_{sw}$ 
for three different angular momenta $l$
[$l=0$ (crosses), $l=1$ (asterisks) and $l=2$ (pluses)].
The depth $V_0$ of  $v_{sw}$
is adjusted so that $1/a_l(k)=0$.
Solid, dotted and dashed lines show fits of the LOCV energies at 
low densities
to Eq.~(\ref{ebpower}) for $l=0$, $l=1$ and $l=2$ (see text for details).
Note that the system with $l=1$ is of theoretical interest but
does not describe a physical system.}
\end{figure}
shows the energy per particle
$E_{B,l}/N$ for a one-component
Bose system, obtained by solving the LOCV equations
for the energetically lowest-lying gas branch, 
as a function of the density $n_B$ for $l=0$ 
(crosses), $l=1$ (asterisks), and $l=2$ (pluses) for the
square well potential, whose depth $V_0$ is adjusted for each $l$ so that the
{\em{energy-dependent}} generalized scattering length $a_l(k)$
diverges, i.e.,
$1/a_l(k) = 0$.
Setting $a_l(k)$ to infinity ensures that the $l$-wave interacting 
Bose system is infinitely strongly interacting over the entire density regime
shown in Fig.~\ref{cap7a}.
Had we instead set the zero-energy scattering length $a_l$ to infinity, the
system would, due to the strong energy-dependence
of $a_l(k)$ [see Fig.~\ref{cap5}], ``effectively'' interact through
a finite scattering length.

Table~\ref{tab1} summarizes the values for $x_l$ and $C_l^G$,
\begin{table}
\begin{tabular}{l|ll l}
$l$ & $x_l$ & $C_l^L$ & $C_l^G$ \\ \hline
0 & 0.00 & $-2.46$ & 13.3 \\
1 & 1.00 & $-3.24$ & 9.22 \\
2 & 2.00 & $-3.30$ & 6.98 
\end{tabular}
\caption{Dimensionless 
parameters $x_l$, $C_l^L$ and $C_l^G$
for $l=0$ to $2$ for a one-component Bose system 
obtained by fitting the LOCV energies $E_{B,l}/N$ for small
densities to the functional form given in Eq.~(\protect\ref{ebpower})
(see text for details).
}
\label{tab1}
\end{table}
which we obtain by performing a  fit of the
LOCV energies $E_{B,l}/N$ for the one-component
Bose system for small densities
to the functional form given in
Eq.~(\ref{ebpower}). 
In particular, 
we find $x_l=l$, which implies that $E_{B,l}/N$ varies as $n_B^{2/3}$,
$n_B^{1/2}$ and $n_B^{1/3}$ for $l = 0$, $1$ and $2$, respectively.
Table~\ref{tab1} uses the superscript ``$G$'' to indicate that the 
proportionality constant is obtained for the 
energetically lowest-lying gas branch. 
The density ranges used in the fit are chosen so
that Eq.~(\ref{ebpower}) describes 
the low-density or universal regime accurately.
Solid, dotted and dashed lines in Fig.~\ref{cap7a} show the results
of these fits
for $l=0$, $1$ and $2$, respectively;
in the low density regime, the lines agree well with the symbols thereby
validating the functional form proposed in Eq.~(\ref{ebpower}).

We repeat the LOCV calculations for the 
energetically highest-lying 
liquid branch of the one-component Bose system. 
By fitting the LOCV energies 
of the liquid branches
for small densities
to Eq.~(\ref{ebpower}), we determine $x_l$ and $C_l$~\cite{footnote4}.
We find the
same $x_l$ 
as for the gas branch but different $C_l$ 
than for the gas branch
(the proportionality constants obtained
for the liquid branch are denoted by $C_l^L$; see Table~\ref{tab1}).
Our values for $x_0$, $C_0^G$ and $C_0^L$ 
agree with those reported in the 
literature~\cite{cowe02,chan04,footnote2}.

\begin{figure}
\vspace{0.2in}
\includegraphics[scale=0.55]{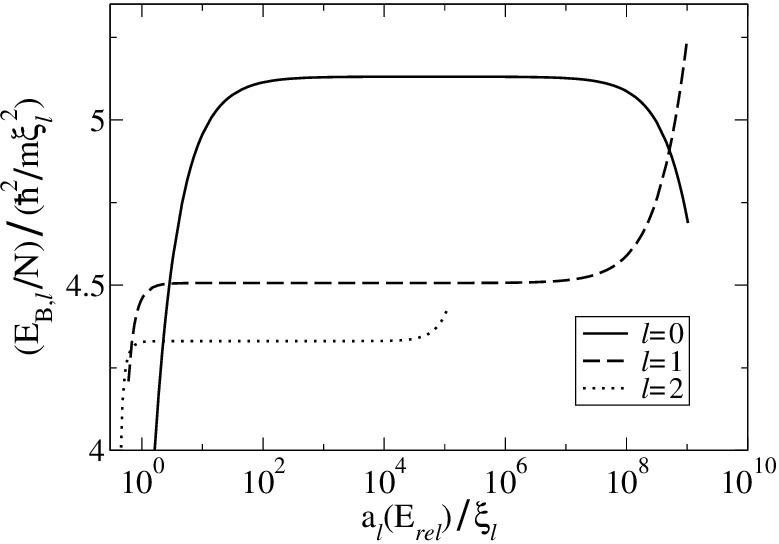}
\caption{\label{cap7} Scaled  energy per particle
$(E_{B,l}/N)/(\hbar^2/m \xi_l^2)$
for the energetically lowest-lying gas branch of
$l$-wave interacting one-component Bose systems 
obtained by solving the LOCV equations
[Eqs.~(\protect\ref{fdfdr_2}) 
through 
(\protect\ref{ebn})] for $v_{sw}$ 
as a function of $a_l(E_{rel})/\xi_l$.
The depth $V_0$ of $v_{sw}$
is adjusted so that
$a_l(E_{rel})=10^{10}R$ for $l=0$ and $l=1$, and $a_l(E_{rel})= 10^{6}R$
for $l=2$.
Note that the system with $l=1$ is of theoretical interest but
does not describe a physical system.
In the regime where the inequality $R \ll
\xi_l \ll a_l(E_{rel})$ is fulfilled, the scaled energy
per particle is constant; this defines the unitary regime.}
\end{figure}

Equation~(\ref{ebpower}) can be rewritten
in terms of the combined length 
$\xi_l$,
\begin{eqnarray}
\label{ebpower2}
\frac{E_{B,l}}{N}=
C_l \Big{(}\frac{4}{3}\pi\Big{)}^{(l/6-2/3)}\frac{\hbar^2}{m
  \xi_l^2},
\end{eqnarray}
where 
\begin{eqnarray}
\label{eqxi}
\xi_l = r_o^{(1-l/4)}R^{l/4}.
\end{eqnarray}
For the $s$-wave case, $\xi_l$ reduces to $r_o$ and the convergence
to the unitary regime can be seen by plotting 
$(E_{B,0}/N)/(\hbar^2/m r_o^2)$ as a function of $a_s/r_o$~\cite{cowe02}.
To investigate the convergence to the unitary regime for higher partial 
waves,
Fig.~\ref{cap7}
shows the energy per particle
$E_{B,l}/N$ for the energetically lowest-lying gas branch
as a function of $a_l(E_{rel})/ \xi_l$ for fixed energy-dependent
scattering lengths $a_l(k)$, i.e.,
for $a_s(k)=10^{10}R$,
$a_p(k)=10^{10}R$, and
$a_d(k)=10^6R$ [the different values of $a_l(k)$ 
are chosen for numerical reasons].
Figure~\ref{cap7} shows that the inequality
\begin{eqnarray}
\label{inequality}
|a_l(E_{rel})| \gg \xi_l \gg R
\end{eqnarray} 
is fulfilled when
$(E_{B,l}/N)/(\hbar^2/m \xi_l^2)$ is constant. 
Note
that this inequality is written in terms of the
energy-dependent scattering length (see above).
We find similar results for the liquid branches for $l=0$ to $2$.
For higher partial waves, we hence use the
inequality given by Eq.~(\ref{inequality}) 
to define the unitary regime.
In the unitary regime,
the energy per particle $E_{B,l}/N$ 
of the Bose system depends only on the
combined length scale $\xi_l$.  
For $s$-wave interacting systems,
we have $\xi_s=r_o$ and $a_s(k)\approx a_s$, and
Eq.~(\ref{inequality}) 
reduces to the well known $s$-wave
unitary condition, i.e., to $|a_s| \gg r_o \gg R$.

We now discuss those regions of Fig.~\ref{cap7}, where 
the energy per particle 
$(E_{B,l}/N)/(\hbar^2/m \xi_l^2)$ for the
one-component Bose system deviates from a constant.
For sufficiently large densities,
the characteristic length $\xi_l$
becomes of the order of the range $R$ of the square well potential.
In this ``high density'' regime 
the system exhibits non-universal behaviors.
In Fig.~\ref{cap7}, e.g., 
the energy-dependent scattering length $a_d(k)$ equals $10^{6}R$;
correspondingly,
$\xi_d$ equals $R$ when
$a_d(k)/ \xi_d = 10^{6}$.  As $a_d(k)/ \xi_d$ approaches $10^6$ from below,
the system becomes non-universal, as indicated 
in Fig.~\ref{cap7} by the non-constant dependence
of the scaled energy per particle on $a_d(k)/\xi_d$.
On the
left side of Fig.~\ref{cap7},
where $a_l(E_{rel})/\xi_l$ becomes of
order $1$,
the ``low density'' end of the unitary regime is reached.
When $a_p(E_{rel})/\xi_p$ equals $10$, e.g., the
interparticle spacing $r_o$ equals
$10^2 a_p(E_{rel})$, i.e., the
system exhibits universal behavior even when the interparticle spacing is
100 times larger than the scattering
length $a_p(E_{rel})$.  
This is in contrast to the $s$-wave case, where the universal
regime requires $|a_s| \gg r_o$. The different behavior of the higher
partial wave systems compared to the $s$-wave system can be understood by
realizing that $\xi_l$ is a combined length, which contains 
both the range $R$
of the two-body potential and the average interparticle spacing $r_o$.
For a given $a_l(E_{rel})/R$, the first inequality in Eq.~(\ref{inequality})
is thus satisfied for larger average interparticle spacings $r_o/R$,
or smaller scaled densities $n_BR^3$, 
as $l$ increases from $0$ to $2$.

In addition to investigating $l$-wave Bose gases interacting through 
the square well potential $v_{sw}$, we consider the van der Waals
potential $v_{vdw}$. For the energetically lowest-lying gas
branch of the one-component 
``$p$-wave Bose'' system we find the same results as for the
square well potential if we replace $R$ in Eqs.~(\ref{ebpower}) to
(\ref{eqxi}) by $\beta_6$. We believe that the same replacement
needs to be done for the liquid branch with $l=1$
and for the liquid and gas branches of $d$-wave interacting bosons,
and that the scaling at unitarity derived above for the square well
potential holds for a wide class of two-body potentials.

Within the LOCV framework, 
the results
obtained for $x_l$, $C_l^G$ and $C_l^L$ 
for the one-component
Bose systems can be applied
readily to the corresponding two-component
system
by scaling the Bose
density appropriately (see Sec.~\ref{sectionII}). 
The resulting parameters $x_l$, $C_l^{L}$
and $C_l^G$ for the two-component
Bose systems are summarized
\begin{table}
\begin{tabular}{l|ll l}
$l$ & $x_l$ & $C_l^L$ & $C_l^G$ \\ \hline
0 & 0.00 & $-1.55$ & 8.40 \\
1 & 1.00 & $-2.29$ & 6.52 \\
2 & 2.00 & $-2.62$ & 5.54 
\end{tabular}
\caption{Dimensionless parameters $x_l$, $C_l^L$ and $C_l^G$
for $l=0$ to $2$ for a two-component Bose system 
(see text for details). 
}
\label{tab2}
\end{table}
in Table~\ref{tab2}.

The energy per particle $E_{F,l}/N$
for $l$-wave interacting two-component Fermi systems
can be obtained from Eq.~(\ref{efn})
using the LOCV solutions 
for the liquid and gas branches
discussed above for $l$-wave interacting one-component Bose 
systems.
In the unitary limit, we find
\begin{equation}
\label{efdwave} 
\frac{E_{F,l}}{N}=A \frac{\hbar^2}{m}n_F^{2/3}
+B_l \frac{\hbar^2}{m} \frac{n_F^{2/3-l/6}}{R^{l/2}},
\end{equation}
where $A=(3/10)(3 \pi^2)^{2/3} \approx 2.87$ and
$B_l=C_l^{G,L} / 2^{2/3-l/6}$
(the $C_l^{G,L}$ are given in Table~\ref{tab1}).
The first term on the right hand side of Eq.~(\ref{efdwave})
equals $E_{FG}$,
and
the second term, which
is obtained from the LOCV solutions,
equals $\lambda_l/2$.
The energy per particle $E_{F,l}/N$ at unitarity
is positive for all densities
for $B_l=C_l^G/2^{2/3-l/6}$. For $B_l=C_l^L/2^{2/3-l/6}$, however,
the energy per particle $E_{F,l}/N$ at unitarity
is negative for $l>0$ 
for small densities, and goes through a minimum
for 
larger densities.
This implies that this branch is
always
mechanically unstable in the dilute limit for $l>0$.

The LOCV treatment for fermions relies heavily on the 
product representation
of the many-body wave function, Eq.~(\ref{eq_wavef}),
which in turn gives rise to the
two terms on the right hand side of Eq.~(\ref{efdwave}).
It is the competition of these two 
energy terms 
that leads to the energy minimum discussed in the previous paragraph.
Future work needs to investigate whether the 
dependence of $E_{F,l}/N$ 
on two length scales as implied by Eq.~(\ref{efdwave}) is correct.
In contrast to the LOCV method,
mean-field treatments predict that 
the energy
at unitarity is proportional
to $E_{FG} |k_F r_{e,l}|$,
where $r_{e,l}$ denotes a range parameter that characterizes
the underlying
two-body potential~\cite{ho05,chen05,iski06}.

\section{conclusion}
\label{sectionV}
This paper investigates Bose and Fermi systems 
using the LOCV method, 
which assumes that three- and higher-order
correlations can be neglected and that the behaviors of the
many-body system are governed by two-body correlations.
This assumption allows the many-body
problem to be reduced to
an effective two-body problem. Besides the
reduced numerical effort, this formalism allows certain aspects of the
many-body physics to be interpreted from a two-body point of view.
Furthermore, it allows parallels between Bose
and Fermi systems to be drawn. 

In agreement with previous studies, we find
that the energy per particle ``corrected'' by the dimer
binding energy, i.e.,
$E_{F,0}/N-E_{dimer}/2$, of dilute two-component $s$-wave
Fermi gases in the whole
crossover regime depends only on the $s$-wave scattering
length
and not on the details of the underlying
two-body potential.
Furthermore, at unitarity the energy per particle is given 
by $E_{F,0}/N=0.46 E_{FG}$. This LOCV
result is in good agreement with the energy per particle obtained
from fixed-node diffusion Monte Carlo calculations, which
predict $E_{F,0}/N=0.42E_{FG}$~\cite{carl03,astr04c,chan04}.
This agreement may be partially 
due to the cancellation of higher-order correlations, and thus 
somewhat fortuitous.
In contrast to Ref.~\cite{gao05}, we find that the liquid branch of 
bosonic helium does not terminate at low densities
but exists down to zero density. 

For higher angular momentum interactions, we determine the 
energy per particle of one- and two-component
Bose systems with infinitely
large scattering lengths. For these
systems, we expect the LOCV formalism to predict
the dimensionless exponent
$x_l$, which determines
the functional dependenc of $E_{B,l}/N$ on 
the range $R$ of the two-body potential
and on the average interparticle spacing $r_o$,
correctly. The values of the proportionality constants
$C_l^G$ and $C_l^L$,
in contrast, may be less accurate.
We use the LOCV energies to generalize the known unitary condition
for $s$-wave interacting systems to systems with finite
angular momentum.
Since
higher angular momentum resonances are
necessarily narrow, leading
to a strong energy-dependence of the scattering strength,
we define the
universal regime
using the
energy-dependent scattering length $a_l(k)$. 
In the unitary regime,
the energy per particle can be written in terms 
of the length $\xi_l$, which is given by a geometric
combination of $r_o$ and $R$. 
The LOCV framework also allows a prediction for the
energy per particle of two-component Fermi gases beyond 
$s$-wave to be made [see Eq.~(\ref{efdwave})]. Although the
functional form of the many-body wave function for
two-component Fermi systems used in this work
may not 
be the best choice,
we speculate that the energy scales derived for strongly interacting
Bose systems are also relevant to Fermi systems.

This work was supported by the NSF through
grant PHY-0555316.
RMK gratefully acknowledges hospitality of the BEC Center,
Trento.


\begin{thebibliography}{10}
\bibitem{footnote3}
{For finite $l$, the term ``unitary regime'' refers to the regime in which the
  observables of the many-body system depend, as discussed in
  Sec.~\protect\ref{sectionIV}, in a simple way on two length scales. Since the
  observables in this regime depend on two length scales, the system
  does not, strictly speaking, exhibit universality. However, because of the
  rather simple dependence on these two length scales, we nevertheless use the
  term ``universal'' to refer to this regime.}
\bibitem{footnote5}
{ The total wave function of the Bose system has to be symmetric with respect
  to the exchange of any two particles, which restricts the angular momentum
  $l$ to even values.}
\bibitem{footnote4}
{For the liquid $l=1$ branch we do not find solutions that obey the boundary
  condition given by Eq.~(\protect\ref{fdfdr_2}). Instead, we determine the
  healing distance $d$ by looking for the minimum of $f'_1$. We find that the
  value of $f'_1$ at the minimum is much smaller than the value of $f'_1$ at
  small $r$.}
\bibitem{footnote2}
{Within the LOCV formalism, the value of $-2.46$ for $C_0^L$ can, as discussed
  in Secs.~\protect\ref{sectionII} and \protect\ref{sectionIII}, be
  straightforwardly converted to obtain the energy per particle $E_{F,0}/N$ of
  the ground state of $s$-wave interacting two-component Fermi gases at
  unitarity, $E_{F,0}/N = 0.46 E_{FG}$~\protect\cite{chan04}. Similarly, the
  value of $13.3$ for $C_0^G$ can be straightforwardly converted to obtain the
  energy per particle $E_{F,0}/N$ of the first excited gas state of $s$-wave
  interacting two-component Fermi gases at unitarity, $E_{F,0}/N = 3.92
  E_{FG}$.}
\end{thebibliography}
\end{document}